\documentstyle[preprint,revtex,eqsecnum]{aps}
\def\PL #1 #2 #3 {Phys. Lett.~{\bf#1} (#2) #3}
\def\NP #1 #2 #3 {Nucl. Phys.~{\bf#1} (#2) #3}
\def\ZP #1 #2 #3 {Z.~Phys.~{\bf#1} (#2) #3}
\def\PR #1 #2 #3 {Phys. Rev.~{\bf#1} (#2) #3}
\def\PRD #1 #2 #3 {Phys. Rev.~D {\bf#1} (#2) #3}
\def\PP #1 #2 #3 {Phys. Rep.~{\bf#1} (#2) #3}
\def\PRL #1 #2 #3 {Phys. Rev.~Lett.~{\bf#1} (#2) #3}
\def\etal {{\it et al}.}

\begin{document}

\hfill\parbox{2in}{\baselineskip14.5pt
{\bf MAD/PH/783}\\
{\bf KEK-TH-370}\\
August 1993}

\vspace{.25in}

\begin{title} Anomalous Higgs Boson Production and Decay
\end{title}
\author{K.~Hagiwara$^{1}$, R.~Szalapski$^{2}$, and
D.~Zeppenfeld$^{2}$}
\begin{instit}
$^1$KEK, Tsukuba, Ibaraki 305, Japan\\
$^2$Department of Physics, University of Wisconsin, Madison, WI 53706, USA
\end{instit}

\thispagestyle{empty}

\begin{abstract}
\nonum\section{abstract}  
\baselineskip14.5pt  
A non-standard symmetry breaking sector may lead to derivative couplings of
the Higgs boson and thereby to anomalous interactions with the electroweak
gauge bosons. In the framework of gauge-invariant effective Lagrangians the
resulting changes in Higgs boson
production and decay mechanisms are related to anomalous triple vector
boson couplings. Using low energy constraints on the dimension-six operators
in the effective Lagrangian, we discuss the size of deviations from SM
predictions which may be expected in Higgs boson production and decay rates.
Large enhancements are allowed {\it e.g.} in the $Z \to H \gamma$ and
$H\to \gamma \gamma$ partial decay widths, leading to $Z\to \gamma\gamma\gamma$
events at LEP.
\end{abstract}

\newpage


In recent years the standard model (SM) of electroweak interactions has
been beautifully confirmed, in particular via the LEP experiments. $Z$
production via $e^+e^-$ annihilation and the bulk
of the $Z$ decay processes mainly probe the couplings of the quarks and
leptons to the $Z$ boson, however. While the gauge theory predictions of these
fermion-$Z$couplings have now been comfirmed at the 1\% level or better, the
bosonic sector of the SM has been tested to a much lesser degree because
present accelerators largely operate below weak boson pair production or Higgs
production threshold.

In order to find out how the ${\rm SU(2)} \otimes {\rm U(1)}$ symmetry of
the SM is broken in nature, an experimental determination is needed of the
interactions between the gauge bosons and the remnants of the
order parameter which gives rise to the spontaneous breaking of the gauge
symmetry, i.e. the Higgs boson in the SM.
In this letter we investigate the phenomenology of models which are relatively
close to the SM in that they posess a (possibly light) Higgs scalar as the
remnant of the SU(2) doublet order parameter.

The anomalous interactions of this doublet field $\Phi$, which posesses the
same quantum numbers as the SM Higgs doublet field, can conveniently be
described by an effective Lagrangian
\begin{eqnarray}\label{Leff}
{\cal L}_{eff} = \sum_i {f_i \over \Lambda^2}{\cal O}_i +
                 \sum_i {f_i^{(8)} \over \Lambda^4}{\cal O}_i^{(8)}+\,...\;\ .
\end{eqnarray}
Here the scale $\Lambda$ may be identified with the typical mass of new
particles associated with the fundamental interactions underlying the symmetry
breaking sector. We here assume that the $W$ and the $Z$ are indeed gauge
bosons of an ${\rm SU(2)} \otimes {\rm U(1)}$ local symmetry. Derivative
couplings of the Higgs, as described by some of the operators in ${\cal
L}_{eff}$, are then related to Higgs--gauge boson ($e.g.$ HVV)
interactions which lead to new phenomena like enhanced $H\to \gamma\gamma$
decay rates or $Z \to H \gamma$ production. The purpose of this letter is to
relate the expected/possible size of such effects to bounds \cite{HISZnew}
derived from present low energy data and to the measurement of anomalous
triple gauge boson vertices (TGV's).

For our analysis it is sufficient to consider the dimension six operators in
${\cal L}_{eff}$ only. This allows a qualitative analysis which is quite model
independent and which is quantitatively correct if
$m_H<<\Lambda$ and $v<<\Lambda$,
where $v$ is the vacuum expectation value of the Higgs doublet field. A
complete analyis of all dimension six operators has been presented by
Buchm\"uller and Wyler\cite{BW}. Here it suffices to consider operators which
can be constructed out of the Higgs field $\Phi$, covariant
derivatives of the Higgs field, $D_\mu \Phi$, and the field strength tensors
$W_{\mu\nu}$ and $B_{\mu\nu}$ of the $W$ and the $B$ gauge fields:
\begin{eqnarray}\label{Wmunu}
[D_\mu,D_\nu] = \hat{B}_{\mu\nu} + \hat{W}_{\mu\nu} =
                i\,{g'\over 2}\,B_{\mu\nu} +
                i\,g\,{\sigma^a\over 2}W^a_{\mu\nu}\; .
\end{eqnarray}
For our discussion of non-standard HVV couplings six such operators need to be
considered, which we call ${\cal O}_{\Phi,1}$, ${\cal O}_{BW}$,
${\cal O}_{W}$, ${\cal O}_{B}$, ${\cal O}_{WW}$, and ${\cal O}_{BB}$.  In
addition the operator ${\cal O}_{\Phi,2}$ contributes via the Higgs boson wave
function renormalization.   They are
given explicitly by\cite{HISZnew}
\begin{eqnarray}\label{operators}
{\cal L}_{eff} & = & \sum_{i=1}^7 {f_i\over \Lambda^2} {\cal O}_i =
{1\over \Lambda^2} \biggl( 
f_{\Phi,1}\; (D_\mu\Phi)^\dagger \Phi\; \Phi^\dagger (D^\mu \Phi)\;
\nonumber \\ & + &
{1 \over 2} f_{\Phi,2}\;  \partial_\mu(\Phi^\dagger \Phi)
\partial^\mu(\Phi^\dagger \Phi)
+\; f_{BW}\; \Phi^\dagger\hat{B}_{\mu\nu}\hat{W}^{\mu\nu} \Phi\;
\nonumber \\
& + & f_W\; (D_\mu\Phi)^\dagger\hat{W}^{\mu\nu}(D_\nu\Phi)\;
+\; f_B\; (D_\mu\Phi)^\dagger\hat{B}^{\mu\nu}(D_\nu\Phi)\;
\nonumber \\
& + & \; f_{WW}\; \Phi^\dagger\hat{W}_{\mu\nu}\hat{W}^{\mu\nu}\Phi\;
+\; f_{BB}\; \Phi^\dagger\hat{B}_{\mu\nu}\hat{B}^{\mu\nu}\Phi
\biggr) \; .
\end{eqnarray}

When the Higgs doublet field is replaced by its v.e.v. one finds that the
operators ${\cal O}_{\Phi,1}$, and ${\cal O}_{BW}$ contribute to gauge boson
self-energies. ${\cal O}_{\Phi,1}$ changes the $Z$ but not the $W$ mass at the
tree level and hence is severely constrained by the measured small value of
$\delta\rho$. ${\cal O}_{BW}$ leads to $W^3$--$B$ mixing and hence contributes
to the $S$ parameter of Peskin and Takeuchi\cite{STU}. A recent
analysis of low energy constraints found~\cite{HISZnew}
\begin{eqnarray}\label{fphi1-fbw}
{f_{\Phi,1}\over \Lambda^2} = (0.11 \pm 0.20)\; {\rm TeV}^{-2} \; , & \; & \;
{f_{BW}\over \Lambda^2} = (1.9 \pm 2.9)\; {\rm TeV}^{-2} \; .
\end{eqnarray}

Similarly the operators ${\cal O}_{W}$ and ${\cal O}_{B}$ give rise to
anomalous TGV's\cite{deR,HISZnew}
\begin{eqnarray}\label{kapanom}
\kappa_\gamma = 1 + (f_B+f_W)\;{m_W^2\over 2\Lambda^2}\; , \;\;
\kappa_Z = 1 + (c^2f_W-s^2f_B)\;
                 {m_Z^2\over 2\Lambda^2}\; , \;\;
g_1^Z = 1 + f_W\;{m_Z^2\over 2\Lambda^2} \; ,
\end{eqnarray}
in the notation of the phenomenological Lagrangian\cite{HPZH}
\begin{equation}
i{\cal L}_{eff}^{WWV} = g_{WWV}\, \left( g_1^V(
W^{\dagger}_{\mu\nu}W^{\mu}-W^{\dagger\, \mu}W_{\mu\nu})V^{\nu} +
\kappa_V\,  W^{\dagger}_{\mu}W_{\nu}V^{\mu\nu} \right)\; , \label{LeffWWV}
\end{equation}
with the overall coupling constants defined as $g_{WW\gamma}=e$ and
$g_{WWZ}= e \cot\theta_W$. Within the SM the couplings are given by
$g_1^Z = g_1^\gamma = \kappa_Z = \kappa_\gamma = 1$.
A direct measurement of the $WW\gamma$ vertex exists from $W\gamma$
production at hadron colliders\cite{UA2} with the result
$\kappa_\gamma = 1^{+2.6}_{-2.2}$ which translates into a measurement of
$f_B+f_W$
\begin{equation}\label{ua2bound}
{f_B + f_W \over \Lambda^2} = 0^{+800}_{-700}\; {\rm TeV}^{-2} \; .
\end{equation}
A more stringent bound can be obtained from an analysis of 1-loop effects of
anomalous TGV's on low energy observables. For $m_H=200$~GeV and a top quark
mass of 140~GeV, one obtains\cite{HISZnew}
\begin{eqnarray} \label{lowedata}
{f_W\over \Lambda^2} = (3\pm 27)\; {\rm TeV}^{-2} \;,\;\;\;\;\;\;\;\;\
{f_B\over \Lambda^2} = (13\pm 19)\; {\rm TeV}^{-2}\; .
\end{eqnarray}
These bounds neglect possible correlations or cancellations between various
new physics contributions, however, and can only be considered as order of
magnitude estimates. The same is true for the coefficients of the operators
${\cal O}_{WW}$ and ${\cal O}_{BB}$, which can only be constrained via their
1-loop contributions to low energy observables prior to Higgs discovery. For
the same assumptions as made for Eq.~(\ref{lowedata}) one finds~\cite{HISZnew}
\begin{eqnarray} \label{lowebbww}
{f_{WW}\over \Lambda^2} = (-7\pm 39)\; {\rm TeV}^{-2}\;,\;\;\;\;\;\;\;\;\
{f_{BB}\over \Lambda^2} = (25\pm 130)\; {\rm TeV}^{-2} \;.
\end{eqnarray}

Allowing for modest correlations/cancellations among the various operators in
their effects on low energy observables, the
coefficients of the four operators ${\cal O}_{W}$, ${\cal O}_{B}$,
${\cal O}_{WW}$, and ${\cal O}_{BB}$ may be as large as $100\; {\rm TeV}^{-2}$,
while for the operators ${\cal O}_{\Phi,1}$ and ${\cal O}_{BW}$ an upper bound
compatible with the low energy data is of order
$|f_i|/\Lambda^2 = 1\; {\rm TeV}^{-2}$. We shall take these values as
illustrative examples in the following to estimate the size of effects in Higgs
boson production and decay. Notice that values $|f_B|/\Lambda^2,\;
|f_W|/\Lambda^2 = 100\; {\rm TeV}^{-2}$ correspond to anomalous TGV's of order
$\kappa_\gamma - 1 = 0.3\; ... \; 0.6$ and hence are exactly in the interesting
range for $W^+W^-$ production experiments at LEP II\cite{LEPWW}. De~R\'ujula
and collaborators \cite{deR}
have argued that it is unnatural to expect a factor of 100
difference between the coefficients of the various operators. We may rather
take the more stringent constraints of Eq.~(\ref{fphi1-fbw}) as an estimate of
the bounds for all
the operators in the effective Lagrangian. Following this more conservative
view we shall consider the case when all coefficients are of order
$|f_i|/\Lambda^2 = 1\; {\rm TeV}^{-2}$ as a second illustrative example for the
consequences of the anomalous HVV couplings.

The $H\gamma\gamma$, $HZZ$, $H\gamma Z$, and $HWW$ couplings follow from the
effective Lagrangian (\ref{Leff}) by making the replacement $\Phi \to
(0,(v+H)/\sqrt{2})^T$. For later use we here list the results in terms of the
HVV effective Lagrangian,
\begin{eqnarray}\nonumber
{\cal L}_{eff}^{HVV} & = & g {m_W \over \Lambda^2} \left\{ -{s^2\left(f_{BB} +
f_{WW} - f_{BW}\right) \over 2} H A_{\mu \nu} A^{\mu \nu} + {2m_W^2 \over
g^2}{f_{\Phi ,1} \over c^2 } H Z_{\mu} Z^{\mu} \right. \\ \nonumber
 &&  \mbox{} + {c^2 f_W + s^2 f_B \over 2 c^2} Z_{\mu \nu} Z^{\mu}
( \partial^{\nu} H) - {s^4f_{BB}+c^4f_{WW}+s^2c^2f_{BW} \over 2 c^2}H
Z_{\mu \nu}
Z^{\mu \nu} \\  \nonumber
&&  \mbox{} + { s\left(f_{W}-f_{B}\right) \over 2 c} A_{\mu \nu} Z^{\mu}
 (\partial^{\nu} H)
+ { s \left(2s^2 f_{BB} -2c^2f_{WW} + (c^2-s^2)f_{BW} \right)\over 2c}
 H A_{\mu \nu} Z^{\mu \nu} \\
 && \left. \mbox{} + {f_W \over 2} \left(W^{+}_{\mu \nu}W^{- \mu} +
W^{-}_{\mu \nu}W^{+ \mu}\right)
(\partial^{\nu} H) - f_{WW} H W^{+}_{\mu \nu} W^{- \mu \nu} \right\} \; ,
\end{eqnarray}
where $
g^2 = e^2 /s^2  =  8 m_W^2 G_F / \sqrt{2}$ .
In addition the operators ${\cal O}_{\Phi,1}$ and ${\cal O}_{\Phi,2}$
renormalize the weak boson masses and the Higgs boson wave function.  By using
the observed masses and couplings the HVV interactions are modified as
\begin{eqnarray}\nonumber
{\cal L}^{HVV}_{ren} & = & g m_W \left[  1 - \left( {f_{\Phi,1}\over 4} +
{f_{\Phi,2}\over 2} \right){v^2 \over \Lambda^2}\right]HW^+_{\mu}W^{-\mu} \\
&& + {1 \over 2} g_Z m_Z \left[ 1 - \left( {f_{\Phi,1} \over
2}+{f_{\Phi,2}\over 2}\right){v^2 \over \Lambda^2}
\right]HZ_{\mu}Z^{\mu} .
\end{eqnarray}
Here the couplings are normalized as $g^2 = \overline{g}^{2}_{W} (0) $ and
$g_Z^2 = \overline{g}^{2}_{Z} (m_Z^2)$ of ref.~\cite{HISZnew}, whose magnitudes
have been precisely measured. Notice that the correction terms in
${\cal L}^{HVV}_{ren}$ need to be considered in addition to the terms given in
Eq.~(10). They are valid only for $|f_{\Phi,1}|v^2/\Lambda^2,
|f_{\Phi,2}|v^2/\Lambda^2 \ll 1$. This condition clearly is satisfied for
$f_{\Phi,1}$ (see Eq.~(4)). The only effect of the operator ${\cal O}_{\Phi,2}$
is a finite wave function renormalization of the Higgs field by a factor
$Z_H^{1\over 2} = 1/\sqrt{1+f_{\Phi,2}v^2/\Lambda^2}$. The phenomenological
consequence is a common rescaling of all Higgs production rates and partial
decay widths by a factor $Z_H$. We shall mostly neglect this overall factor
(by setting $f_{\Phi,2}=0$) when considering the nonstandard contributions to
$H\to VV$ decay rates and $Z \to HV$ decay, which are induced by the
anomalous HVV couplings.


{\bf Higgs Decays}.  Higgs decays into $\gamma\gamma$, $Z\gamma$, $ZZ$ and
$W^+W^-$ are affected by non-standard interactions from dimension-six
operators.
In the SM the decay $H \rightarrow \gamma \gamma$ occurs at the
one-loop level.  All massive particles with non-zero electromagnetic charge
run through the loop, the most important contributions arising from the
top quark and the $W$ boson. However, dimension-six operators contribute
at
the tree level and can therefore lead to large deviations from SM
expectations~\cite{randall}. The total width for this process is given by
\begin{equation}\label{h2gam}
\Gamma (H \rightarrow \gamma \gamma) = {\alpha s^2 m_W^2 m_H^3
\over 4 } \left| {f_{BB} + f_{WW} - f_{BW} \over \Lambda^2} + { \alpha \over
8 \pi s^2 m_W^2} I \right|^2 \; ,
\end{equation}
with $s = \sin \theta_W$.
The SM contribution is parametrized by the complex-valued function
$I = \sum_{i}N_{ci} e_i^2 F_i $ where $N_{ci}$ is the color multiplicity of
particle i and $e_i$ is its charge.  The functions $F_i$ are given
explicitly in Ref.~\cite{hunter}.

In the SM the branching fraction for $H \rightarrow \gamma \gamma$ is ${\cal
O} (10^{-3})$.  Although it is a rare decay it is the primary search mode for
an intermediate-mass Higgs boson at hadron colliders where the $H \rightarrow b
\overline{b}$ signal is swamped by QCD backgrounds, but $H \rightarrow WW, ZZ$
is not kinematically allowed.  For maximal values of
$f_{BB}$ and $f_{WW}$ each $( \sim {\cal O}(100\;{\rm TeV}^{-2}))$,
and assuming no large cancellations, the width for $H\rightarrow\gamma\gamma$
receives an enhancement by a factor $10^4$ and, below the $H \rightarrow WW^*$
``threshold'', is the primary decay mode, see fig.~1a.  The enhanced width is
shown by the dashed line, and the SM prediction is given by the solid line.
For a heavier Higgs it
may still be a competitive process with a branching fraction of a few percent.
Hence, the search for an intermediate mass Higgs boson at hadron
colliders would be greatly facilitated.  Note, however, that a large
$H \gamma \gamma$ coupling and large TGV's occur in orthogonal
directions in the $f_i$ parameter space.  Hence a large rate for $H
\rightarrow \gamma \gamma$ does not neccesarily imply large TGV's.

For ``natural'' values of the $f_i$ ( $f_i/\Lambda^2 =
1{\rm TeV}^{-2}$ for
all $f_i$) the SM one-loop contribution and the
dimension-six contributions are comparable; significant interference is, in
principle, possible. Generically one might expect a change in the rate for $H
\rightarrow \gamma \gamma$ by a factor of $2$ or so.
The dotted line displays destructive interference while
the double-dotted line shows constructive interference.

The process $H \rightarrow \gamma Z$ is very similar to the process
$H \rightarrow \gamma \gamma$; the SM contribution is a one-loop process, but
the dimension-six contribution occurs at the tree level.  The total width for
this process is given by
\begin{eqnarray}\label{hgz}
\Gamma ( H \rightarrow \gamma Z) &=& {\alpha ( m_H^2-m_Z^2)^3 m_Z^2 \over 16
m_H^3} \\ \nonumber &&
\left| { f_W - f_B + 4 s^2 f_{BB} - 4 c^2 f_{WW} + 2(c^2-s^2)f_{BW}
\over \Lambda^2} + {\alpha \over 2 \pi sc m_Z^2 } A \right| ^2\; .
\end{eqnarray}
The SM contribution is parametrized by the complex-valued function
$ A = A_F + A_W $ which is
given explicitly in Ref.~\cite{hunter}.

Setting ${ f_i/ \Lambda^2} = 100 {\rm TeV}^{-2}$ for all $f_i$ except $f_{BW}$
then the width for $H \rightarrow \gamma
Z$ is enhanced by a factor of $10^3$.  See the dashed line in fig. 1b.
The resulting branching fraction is around $10\%$ when the decay is
kinematically allowed.
Hence $H \rightarrow \gamma Z$ would be a complementary channel to
$H \rightarrow \gamma \gamma $ for $ 100 {\rm GeV} \leq m_H \leq 140 {\rm
GeV}$.
For a heavier Higgs boson both  $H \rightarrow \gamma Z$ and
$H \rightarrow \gamma \gamma$ would complement $H \rightarrow WW, ZZ$.
The appearence of $f_{BB}$ and $f_{WW}$ in Eq.~(\ref{hgz}) means that one can
have an enhanced $H \rightarrow \gamma Z$ rate without large anomalous
TGV's.  However, the appearance of $f_B$ and $f_W$ in Eq.~(\ref{hgz})
implies that, barring large accidental cancellations, measurable values for
$\Delta\kappa_\gamma$, $\Delta\kappa_Z$ and $g_1^Z$ imply a strongly enhanced
$H \rightarrow \gamma Z$ rate. E.g. a value of $\kappa_\gamma\approx 0.2$, at
the limit of observability in $W^+W^-$ production at LEP II\cite{LEPWW},
implies values of $(f_B+f_W)/\Lambda^2$ in the vicinity of the values used for
the dashed line in fig.~1b.

For more ``natural'' values of the $f_i$ ($f_i/\Lambda^2 \sim {\cal
O}(1 {\rm TeV}^{-2})$) the SM one-loop contribution and the
dimension-six contribution are comparable and interference is probable.  One
might expect a change in this rate by a factor of two or so, and this decay
remains a rare process. See the dotted and double-dotted lines in fig. 1b.
As such this channel is not likely to be instrumental
in the discovery of the Higgs boson, but it does serve as an important
precision
test of the SM.  Along with a measurement of the $H \rightarrow \gamma \gamma$
rate and the measurement of anomalous TGV's a measurement of the $H
\rightarrow \gamma Z$ rate places some important restrictions on the allowed
directions in the $f_i$ parameter space.  As in the
case of $H \rightarrow \gamma \gamma$ there is a small region of parameter
space corresponding to maximal destructive interference, in which case this
mode would be unobservable.

The decays $H \to ZZ$ and $H \to WW$ occur at tree level, and hence they are
affected significantly only when we allow larger magnitudes for the
coefficients of dimension-six operators.
Separating the result into longitudinal and transverse contributions
the decay width for the process $H \rightarrow ZZ$ is given by
\begin{eqnarray}\label{zztrans}
\Gamma(H \rightarrow Z_T Z_T) &=& {g^2 \over 128 \pi} {m_H^3 \over m_W^2 }
\sqrt{1-x_z}  {1 \over 2 } \left[ x_z(1+{\cal C}_{ZZ}) + {\cal D}_{ZZ}
\right]^2\; ,
\end{eqnarray}
and
\begin{eqnarray}\label{zzlong}
\Gamma(H \rightarrow Z_L Z_L) &=& {g^2 \over 128 \pi} {m_H^3 \over m_W^2 }
\sqrt{1-x_z}  {1 \over 4 } \left[ (2-x_z)(1+{\cal C}_{ZZ}) + {\cal D}_{ZZ}
\right]^2 \; ,
\end{eqnarray}
where
\begin{eqnarray}\nonumber
{\cal C}_{ZZ} &=& (f_{\Phi,1} - f_{\Phi,2}){v^2\over 2\Lambda^2}
-2 {m_Z^2 \over \Lambda^2}
\left[ c^4 f_{WW}+s^2 c^2 f_{BW}  +s^4 f_{BB} \right], \\ \nonumber
{\cal D}_{ZZ} &=& 2 {m_Z^2 \over \Lambda^2} \left[ 2c^4 f_{WW}
+ 2 s^2 c^2 f_{BW} +2s^4 f_{BB}- s^2f_B - c^2f_W \right] \; ,
\end{eqnarray}
and $x_z = 4 m_Z^2 / m_H^2$.

The width for $H \rightarrow WW$ is very similar.
\begin{eqnarray}\label{wwtrans}
\Gamma(H \rightarrow W_T W_T) &=& {g^2 \over 64 \pi} {m_H^3 \over m_W^2 }
\sqrt{1-x_w} {1 \over 2}\left[ x_w(1+{\cal C}_{WW}) + {\cal D}_{WW}
\right]^2 \; ,
\end{eqnarray}
and
\begin{eqnarray}\label{wwlong}
\Gamma(H \rightarrow W_LW_L) &=& {g^2 \over 64 \pi} {m_H^3 \over m_W^2 }
\sqrt{1-x_w}  {1 \over 4 } \left[ (2-x_w)(1+{\cal C}_{WW}) + {\cal D}_{WW}
\right]^2 \; ,
\end{eqnarray}
where
\begin{eqnarray}\nonumber
{\cal C}_{WW} &=& -(f_{\Phi,1} + 2f_{\Phi,2}) {v^2\over 4\Lambda^2}
 -2 {m_W^2 \over \Lambda^2}f_{WW} ,\\ \nonumber
{\cal D}_{WW} &=& 2 {m_W^2 \over \Lambda^2} (2 f_{WW} - f_{W} ),
\end{eqnarray}
and $x_w = 4 m_W^2 / m_H^2$.

When anomalous contributions vanish, i.e. $f_i/\Lambda^2 \rightarrow 0$ for
all $f_i$, then ${\cal C}_{VV}, {\cal D}_{VV} \rightarrow 0$ for $V = W, Z$
and the SM value for $\Gamma(H \rightarrow VV)$ is recovered.  If
${\cal D}_{VV} = 0$ then the $H \rightarrow VV$ decay width is enhanced by an
overall factor $(1+{\cal C}_{VV})^2$.  However, a non-zero value of
${\cal D}_{VV}$ will change the ratio of longitudinally polarized and
transversely polarized vector bosons. It is possible to construct a scenario
where ${\cal C}_{WW}, {\cal C}_{ZZ} \neq 0$ with ${\cal D}_{WW},
{\cal D}_{ZZ} = 0$, but only in a very restricted region of parameter space.
Hence, one should in general expect that a change in the overall rate will
be accompanied
by a change in the ratio of longitudinal and transverse polarizations.

The size of expected deviations in the $H\to WW$ and $H\to ZZ$ partial widths
is shown in figs.~1c and 1d.
For ``natural'' values of the $f_i$ ($f_i/\Lambda^2 \sim {\cal O}
(1{\rm TeV}^{-2})$)  we do not
expect measurable changes in these decay rates; the dotted and double-dotted
lines are nearly degenerate with the solid SM curves.  However, if
the $f_i$ approach their phenomenological bounds then modest effects are
likely to be seen. In this case destructive interference effects between SM
and new physics contributions are possible, independently in the $H\to WW$ and
$H\to ZZ$ decay modes. Hence the ratio of $WW$ vs. $ZZ$ Higgs signals at
hadron supercolliders  may be drastically altered.



{\bf Z decays involving the Higgs boson}.  The new interactions from
dimension-six operators can significantly effect the $Z$ decay branching
fractions $Z \to HZ^{*} \to H f \overline{f} $, $Z \to H \gamma$ and
$Z \to H \gamma \to \gamma \gamma \gamma$.
We present the differential decay rate for
$Z \rightarrow HZ^* \rightarrow Hf\overline{f}$ in terms of the individual
particle momenta.  Denote the four-momentum of the initial-state
$Z^0$ by $p_Z$.
$p_H$ denotes the four-momentum of the Higgs boson and $E_H$ is its energy in
the rest frame of the initial-state $Z^0$.
Finally, $q_1$ and $q_2$ denote the fermion and anti-fermion four-momenta, and
$E_f$ is the energy of the fermion.  Then
\begin{eqnarray}\nonumber
{d\Gamma \over dE_f dE_H} & = &
                {g^4 \over 24 (2\pi)^3 c^6}{m_W^2 \over m_Z^3}
        { g_V^2 + g_A^2 \over \left[ (p_Z - p_H)^2 - m_Z^2 \right]^2 + (m_Z
        \Gamma_Z)^2 } \\ \nonumber &&
        \Bigg\{ 2 {\cal G}_1^2 q_1 \cdot q_2
        + ( q_1 \cdot p_H q_2 \cdot p_H - m_H^2 q_1 \cdot q_2 ) \\ &&
        \left[ { {\cal G}_1^2 \over m_Z^2 }
        + 2 {\cal G}_1{\cal G}_2 \left( 1 - { p_Z \cdot p_H \over m_Z^2
}\right)
        -{\cal G}_2^2 \left( m_H^2 - { (p_Z \cdot p_H) ^2 \over m_Z^2} \right)
        \right] \Bigg\},
\end{eqnarray}
where
\begin{eqnarray}\nonumber
{\cal G}_1 &=& -2\left( 1+(f_{\Phi,1}-f_{\Phi,2}){v^2\over 2\Lambda^2}\right)
        +m_H^2 {s^2f_B+c^2f_W \over \Lambda^2 }\\ && \mbox{}
  + 4 (m_Z^2-p_Z \cdot p_H){ s^4f_{BB}+s^2c^2f_{BW} +c^4f_{WW}\over
        \Lambda^2 }  \; , \\
{\cal G}_2 &=& 2 { s^2f_B + c^2 f_W \over \Lambda^2 } -
        4 { s^4f_{BB}+s^2c^2f_{BW}+c^4f_{WW} \over \Lambda^2 },
\end{eqnarray}
and $g_V = {1 \over 2} T_3 - s^2 Q$ and $g_A=- {1 \over 2} T_3$ are the vector
and axial-vector couplings of the fermion.

For ``natural'' values of all $f_i$( $f_i \sim {\cal O}(1{\rm TeV}^{-2})$)
new physics effects are negligible.  For phenomenologically allowed values of
$f_i \sim {\cal O}(100{\rm TeV}^{-2})$ effects remain modest (see the dashed
lines in fig.~2a.) but are important with regard to the ongoing Higgs boson
search at LEP.
Assuming SM couplings the Higgs boson is currently constrained to be heavier
than approximately 60 GeV by searching in the channel
$Z \rightarrow HZ^* \rightarrow Hf\overline{f}$. Contributions from
dimension-six interactions can weaken this bound significantly.

The decay $Z \rightarrow H \gamma$ occurs in the SM at the one-loop level; the
dimension-six contribution occurs at the  tree level and has been considered
previously by several authors\cite{randall}.  The combined width
for this process is
\begin{eqnarray}\nonumber\label{hzg}
\Gamma(Z \rightarrow H \gamma) &=& {\alpha \over 96 m_Z}
\left(m_Z^2-m_H^2\right)^3  \\ \label{zhg} &&
\left| {f_W - f_B + 4s^2f_{BB} -4c^2f_{WW} +
2(c^2-s^2)f_{BW} \over \Lambda^2} + {\alpha \over 2 \pi sc m_Z^2}A
\right|^2.
\end{eqnarray}
The SM contribution is parametrized by the complex-valued function $A$, which
is given explicitly in Ref.~\cite{hunter}.
If all $f_i$ are of order $f_i/\Lambda^2 \sim {\cal O} (1{\rm TeV}^{-2})$ then
the SM and
the dimension-six contributions are comparable and, unless there is maximal
destructive interference between the SM and the new contributions we do
not expect a large change in the rate for $Z \rightarrow H \gamma$.
However, the contributions from dimension-six operators completely dominate
this partial decay
rate of the $Z$ if the $f_i$ are close to their phenomenological limits (see
fig.~2b), and there is a large rate
enhancement.  The search for the Higgs boson in this channel
actually provides a constraint on the linear combination of the $f_i$ appearing
in Eq.~(\ref{hzg}) (providing $m_H < m_Z$) since, as can be seen from the
topmost curve (dashed line) in fig. 2b, for
some allowed values of the $f_i$ a light Higgs boson should have already been
discovered. Notice that $\Gamma(Z \rightarrow Hf\overline{f})$ and
$\Gamma(Z \rightarrow H\gamma)$ involve different linear combinations of the
$f_i$, hence one process may be affected but not the other.  In particular a
reduced rate in the $Z \rightarrow Hf\overline{f}$ channel does not
neccesarily imply an enhanced rate in the $Z \rightarrow H\gamma$ rate, though
an enhancement is likely. Therefore, new physics of the type discussed here
can weaken the lower limit on the mass of the Higgs boson obtained at LEP.

The sequential decay $Z \to H \gamma \to \gamma \gamma \gamma$ is significant
if both $Z \to H \gamma$ and $H \to \gamma \gamma$ branching
fractions are enhanced by the new interactions.
In the SM the decay $Z \to \gamma \gamma \gamma$ occurs at the one loop level.
The contribution due to fermions in the loop has been calculated
\cite{fermions} and, for a heavy top quark, the contribution is
0.7 eV.  The contribution due to W bosons in the loop has also been
calculated \cite{bosons}.  This contribution is found to be smaller by a
factor of 35.  Interference effects between bosonic and fermionic loops have
not been calculated.  We ignore the contribution from W bosons.
$\Gamma(Z \to \gamma \gamma \gamma)$ also receives a contribution via $Z \to H
\gamma \to \gamma \gamma \gamma$.  In the purely SM scenario both the $H Z
\gamma$ vertex and the $H \gamma \gamma$ vertex are generated at one loop,
hence the SM contribution is non-negligible only if $Z \to H \gamma$ is
kinematically allowed.

With the inclusion of dimension-six effects both the $HZ \gamma$ vertex and
the $H \gamma \gamma$ vertex may be enhanced.  In this scenario the
process $Z \to H \gamma \to \gamma \gamma \gamma$ may be important even for a
virtual Higgs boson.  Because this process then involves the product of two
dimension-six operators, our calculation via virtual Higgs boson exchange
should be regarded as an estimate of the possible dimension-eight
$Z\gamma\gamma\gamma$ vertex.
The results are sumarised by fig. 3.  The SM contribution is too small to be
interesting.  For small dimension-six contributions ($f_i/\Lambda^2 \sim 1
{\rm TeV}^{-2}$)  the effects are also very small.

The situation is dramatically different for large dimension-six effects
($f_i/\Lambda^2 \sim 100 {\rm TeV}^{-2}$).   Based upon fig. 3 the search
for $Z \to \gamma \gamma \gamma$ events becomes promising for a light Higgs
boson.  There is even some
hope for events in this channel for $m_H > m_Z$ if the Higgs boson is not too
much heavier than $m_Z$.

{\bf Summary}.
New physics in the electroweak bosonic sector may be described by an effective
Lagrangian of dimension-six operators.  The coefficients of some of these
operators are severely constrained by low-energy data ($f_i/\Lambda^2 <
{\cal O}(1{\rm TeV}^{-2})$), while others may be as large as
$f_i/\Lambda^2 \sim {\cal O}(100{\rm TeV}^{-2})$.
Actually, allowing for arbitrary cancellations amongst the full set of
operators no stringent and rigorous bounds exist on any of them.

In the pessimistic scenario ($f_i/\Lambda^2 < {\cal O}(1{\rm TeV}^{-2})$ for
all
of the operators) one does not expect to see deviations from the SM
predictions for TGV's.  Furthermore, one does not expect
to observe changes in SM processes which occur at the tree level.  However,
proceeses which occur at one-loop in the SM but have tree-level dimension-six
contributions could differ from their SM expectations appreciably.  This is
demonstrated in Fig. 4b where the various Higgs boson branching fractions are
compared. While most of the branching fractions are indistinguishable from
their SM values, the $H\to \gamma\gamma$ and $H\to Z\gamma$ rates are strongly
affected, which would have important consequences for intermediate-mass Higgs
boson searches at hadron colliders.

Huge effects on Higgs phenomenology are possible if large dimension six
contributions close to their present phenomenological
low energy bounds are realized in nature .
 This is demonstrated by the Higgs branching ratios of fig.
4a: an intermediate mass Higgs might predominantly decay into two photons.
Actually large effects like the ones in fig.4a should be expected if anomalous
TGV's are large enough to be observed in $W^+W^-$
production at LEP II. Thus the search for a light Higgs at LEP I may have
important consequences for vector boson pair production at higher energies.

{\bf Acknowledgements}
Discussions with T.~Hatsukano and S.~Ishihara are gratefully acknowledged.
This research was supported in part by the University of Wisconsin Research
Committee with funds granted by the Wisconsin Alumni Research Foundation,
by the U.~S.~Department of Energy under contract No.~DE-AC02-76ER00881,
and by the Texas National Research Laboratory Commission under Grants
No.~RGFY9273 and FCFY9212.

\figure{Higgs boson decay widths for the channels a) $H \to \gamma \gamma$,
b)$H \to Z\gamma$,
c)$H \to WW$, d)$H \to ZZ$.  The solid line is purely SM, dots:
$f_i/\Lambda^2 = 1{\rm TeV}^{-2}$ for all six operators,
double dots: $f_i/\Lambda^2 = -1{\rm TeV}^{-2}$
for all six operators, dashes: $f_i/\Lambda^2 = 1{\rm TeV}^{-2}$ for $f_{BW}$
and $f_{\Phi,1}$, while $f_i/\Lambda^2 = 100{\rm TeV}^{-2}$ for $f_{BB}$,
$f_{WW}$,
$f_B$, and $f_W$, and long-dash short-dash: $f_i/\Lambda^2 = 1{\rm TeV}^{-2}$
for $f_{BW}$ and $ f_{\Phi,1}$, $f_i/\Lambda^2 = -100{\rm TeV}^{-2}$ for
$f_{BB}$, $f_{WW}$, $f_B$, and $f_W$. \label{higgsdecays} }

\figure{Partial $Z$ decay widths with a Higgs boson in the final state.
a)$Z \to H f\overline{f}$ summed over all kinematically allowed SM fermions,
b)$Z \to H \gamma$.  The various lines are for the same linear combinations of
dimension-six operators as in fig.~\ref{higgsdecays}: the solid line is purely
SM, the dotted lines describe the effect of ``phenomenologically allowed''
coefficients $f_i/\Lambda^2$ while the dashed lines show examples of
``natural'' values of these coefficients.  \label{zdecays} }

\figure{$\Gamma(Z \to \gamma \gamma \gamma)$ in various scenarios.
The solid line is purely SM, dots:
$f_i/\Lambda^2 = 1{\rm TeV}^{-2}$ for all six operators,
double dots: $f_i/\Lambda^2 = -1{\rm TeV}^{-2}$
for all six operators and dashes: $f_i/\Lambda^2 = 1{\rm TeV}^{-2}$ for $f_{BW}
$
and  $f_{\Phi,1}$, $f_i/\Lambda^2 = 100{\rm TeV}^{-2}$ for $f_{BB}$, $f_{WW}$,
$f_B$ and $f_W$.  The remaining curve (dash double dot) is the SM width for $Z
\to H \gamma$, included for reference. \label{zto3gam} }

\figure{Higgs boson branching fractions for two choices of dimension-six
operators:  a) $f_i/\Lambda^2 = 1{\rm TeV}^{-2}$
for $f_{BW}$ and
$f_{\Phi,1}$, $f_i/\Lambda^2 = 100{\rm TeV}^{-2}$ for $f_{BB}$,
$f_{WW}$, $f_B$ and $f_W$, b)$f_i/\Lambda^2 = 1{\rm TeV}^{-2}$ for all six
operators. \label{bf} }


\begin{references}

\bibitem{HISZnew}
K.~Hagiwara, 
S.~Ishihara, R.~Szalapski, and D.~Zeppenfeld,
preprint MAD/PH/737 (1993), Phys. Rev. {\bf D48}, in press.

\bibitem{BW}
W.~Buchm\"uller and D.~Wyler, Nucl.\ Phys.\ {\bf B268} (1986) 621.

\bibitem{STU}
M.~E.~Peskin and T.~Takeuchi, \PRL 65 1990 964 .

\bibitem{deR}
A.~De~R\'ujula, M.~B.~Gavela, P.~Hernandez, E.~Mass\'o,
Nucl.\ Phys. {\bf B384} (1992) 3.   

\bibitem{HPZH}
K. Hagiwara, K.~Hikasa, R.D.~Peccei, D.~Zeppenfeld, \NP B282 1987 253 .

\bibitem{UA2}
UA2 collaboration, J.~Alitti \etal , Phys. Lett. {\bf B277} (1992) 194;
H.~Aihara, talk given at the conference on {\it Physics in Collisions},
Heidelberg, June 1993.

\bibitem{LEPWW}
G.~Barbiellini \etal, in ``Physics at LEP,'' ed.\ by J.~Ellis and
R.~D.~Peccei, report CERN 86-02, Vol.~2, p.~1;
M.~Davier \etal, ECFA Workshop on LEP 200, ed.\ by A.~Bohm and W.~Hoogland,
CERN report CERN 87-08, Vol.~I, p.~120.

\bibitem{randall}
D.~D\"usedau and J.~Wudka, \PL 180B 1986 290 ;
L.~Randall and N.~Rius, \PL B309 1993 365 .


\bibitem{hunter}
J.~F.~Gunion, H.~E.~Haber, G.L.~Kane, and S.~Dawson, The Higgs Hunter's Guide,
Addison-Wesly, 1990.

\bibitem{fermions}
J.~J.~van~der~Bij and E.~W.~N.~Glover, Nucl. Phys. {\bf B313} (1989) 237;
See also J.~J.~van~der~Bij and E.~W.~N.~Glover in,`` Z Physics at LEP I'',
ed.\ by  G.~Altarelli, R.~Kleiss, and C.~Verzegnassi, CERN report CERN 89-08,
Vol. 2, p. 30;
M.L.~Laursen, K.O.~Mikaelian and A.~Samuel, Phys. Rev. {\bf D23} (1981) 2795.

\bibitem{bosons}
M.~Baillargeon and F.~Boudjema , Phys. Lett. {\bf B272} (1991) 158;
Fang-xiao~Dong, Xiang-dong~Jiang, Xian-jian~Zhou, Phys. Rev. {\bf D46} (1992)
5074.

\end{references}
\end{document}